\documentclass[12pt]{iopart}
\usepackage{cite}
\usepackage{mathrsfs}
\usepackage{bm}
\usepackage{graphicx} 
\usepackage{hyperref}

\begin{document}

\title[Geometrical Lagrangian  statistics of heavy impurities in drift-wave turbulence]{Multiscale geometrical Lagrangian  statistics of heavy impurities in drift-wave turbulence}

\author{Zetao Lin$^1$, Benjamin Kadoch$^2$,\\
Sadruddin Benkadda$^3$ and Kai Schneider$^1$}

\address{$^1$ Aix-Marseille Universit\'e, CNRS, I2M, UMR 7373, 13331 Marseille, France}
\address{$^2$ Aix-Marseille Universit\'e, CNRS, IUSTI, UMR 7343, 13453 Marseille, France}
\address{$^3$ Aix-Marseille Universit\'e, CNRS, PIIM, UMR 7345, 13397 Marseille, France}
\ead{zetao.lin@etu.univ-amu.fr}
\vspace{10pt}
\begin{indented}
  \item[] March 2025  
\end{indented}

\begin{abstract}
We investigate the behavior of heavy impurities in edge plasma turbulence by analyzing their trajectories using the Hasegawa--Wakatani model. Through direct numerical simulations, we track ensembles of charged impurity particles over hundreds of eddy turnover times within statistically steady turbulent flows. Assuming that heavy impurities lag behind the flow,  a novel derivation of relaxation time of heavy impurities is proposed. Our results reveal that  heavy impurities can cluster within turbulence.  We  provide multiscale geometrical Lagrangian  statistics of heavy impurities trajectories. To quantify directional changes, we analyze the scale-dependent curvature angle, along with the influence of the Stokes number on the mean curvature angles and the probability distribution function of curvature angles.

\end{abstract}

\section{Introduction}

Observations in tokamaks have shown that impurity transport often exceeds what neoclassical theory predicts, with turbulence being regarded as the origin of this anomalous transport. Drift-waves are believed to have a significant impact on the dynamics and transport characteristics of edge turbulence in tokamaks, as discussed in \cite{Futatani2008b} and related literature. At the edge of the plasma, where temperatures are low and collision rates are high, resistivity becomes a crucial factor. The Hasegawa--Wakatani model is a two-field model that incorporates key aspects of turbulent transport through resistive drift-waves. Despite its underlying simplifying assumptions, it contains the basic elements to investigate transport, such as a broad range of turbulent fluctuations and the spontaneous formation of coherent structures. In this study, the two-dimensional slab geometry version of the model is chosen as a paradigm for drift-wave turbulence in the plasma-edge region.  

The transport of impurities can be analyzed through two complementary approaches. The first approach involves a passive scalar method, where impurity density within a fluid element is assessed (see, for instance, \cite{Futatani2008a, Futatani2009a}). This method relies on an advection-diffusion equation for a passive scalar. Alternatively, the transport properties can be investigated by tracking individual marked particles. This perspective involves studying the trajectories of multiple tracer particles through numerical simulations, solving equations describing the motion of these particles within specific plasma flow velocity fields, such as the $\bm {E} \times \bm{B}$ field.

In fusion plasma research, previous studies have investigated passive flow tracers in edge plasma using the Hasegawa--Wakatani model, without considering inertial effects \cite{Futatani2008b, Futatani2008a, Futatani2009a}. While this approximation works well for light impurities, it may become inadequate for future fusion devices such as ITER, where heavy impurities, like tungsten, are expected to be present. Tungsten impurity, due to their significant inertia, may exhibit behaviors that are not captured by the existing models. In the work of Priego \textit{et al.} \cite{Priego2005}, inertial effects were considered, employing a fluid model for impurities in which the impurity velocity was described as the sum of the $\bm{E} \times \bm{B}$ and polarization drifts. The polarization drift, which accounts for impurity particle inertia, serves as a higher-order correction to the $\bm{E} 	\times \bm{B}$ drift velocity and introduces compressible effects.

In contrast to fluid modeling, our approach tracks each impurity particle individually. In the work of Priego \textit{et al.} \cite{Priego2005}, the inertial effect is linked to polarization drift, which only appears in fusion plasma, not in hydrodynamics. Here we investigate the inertial effect that is linked to drag force, which is well established in fluid dynamics. Our contribution is the novel  derivation of the relaxation time of particles in plasma environment. We  hypothesize that heavy impurity particles may ``lag behind'' the plasma flow velocity due to their inertia \cite{Lin2024a, Lin2025}. The preferential concentration of inertial particles in regions of low vorticity is a well-established phenomenon in fluid turbulence \cite{Maxey1987}.  In our study of fusion plasmas, we observe a similar trend: heavy impurity particles tend to cluster in low-vorticity regions, where coherent vortices significantly influence their spatial distribution. In our earlier studies \cite{Lin2024a, Lin2025}, the relaxation time was not explicitly defined; instead, it was varied over several orders of magnitude to investigate the influence of inertia on particle dynamics. The key advancement in the present work lies in deriving an exact expression for the relaxation time, offering a more robust and precise foundation for analyzing these phenomena.

Complex dynamics emerge from ensembles of interacting elements. Individual particle trajectories hold insights beyond overall ensemble behavior, revealing new mechanistic aspects of the system. A particle's trajectory forms a curve in 2D or 3D, and intrinsic geometric analysis requires curvature determination. In fact, the shape of a smooth curve with non vanishing curvature is completely described by curvature and torsion (in 3D), the so-called Frenet–Serret formula.  In turbulent flow, spirals are prototypical curves, and their association with vortex-like motion suggests that curvature provides information on particle inertia.  In fluid mechanics, instantaneous measurements of turbulence curvature in academic flows have been studied \cite{Braun2006, Xu2007, Kadoch2011}. A scale-dependent curvature measure was proposed in \cite{Burov2013, Bos2015}, demonstrating how the turbulent flow's multiscale nature can be understood through directional changes. For the first time in fusion plasma, our study  provides analyses of  multiscale geometrical Lagrangian statistics for heavy impurities in drift-wave turbulence. Here the scale-dependent curvature angle is used to quantify the  directional changes of heavy impurities. The impact of the Stokes number on the mean curvature angles and probability distribution function of curvature angles on different time scale is analyzed.

The remainder of the paper is organized as follows.
Section~\ref{sec: Models and Method} presents briefly the 
Hasegawa--Wakatani model and the governing model equations for charged heavy impurity particles with a detailed derivation of the parameters. Multiscale geometrical statistics considering the scale-dependent curvature angle of the impurity trajectories are likewise introduced.
Numerical results are then presented in section~\ref{sec:results} and conclusions are drawn in section~\ref{sec:conclusion}.

\section{\label{sec: Models and Method}Models for simulation and analysis method}

\subsection{Hasegawa--Wakatani model}

We perform direct numerical simulations (DNS) of  plasma edge turbulence driven by drift-wave instability \cite{Hasegawa1983} using a pseudo-spectral code \cite{Kadoch2022}. Our focus in this study is on the two-dimensional slab geometry of the Hasegawa--Wakatani model (HW model), see, \textit{e.g.}, \cite{Bos2010,Kadoch2022}. The standard form of the 2D HW model is given by,
\begin{equation}
\left(\frac{\partial}{\partial t}- \mu_{\nu} \nabla^{2}\right) \nabla^{2} \phi=\left[\nabla^{2} \phi, \phi\right]+c(\phi-n),
\label{eq:phi}
\end{equation}

\begin{equation}
\left(\frac{\partial}{\partial t}-\mu_{D} \nabla^{2}\right) n=[n, \phi]-\Gamma \frac{\partial \phi}{\partial y}+c(\phi-n),
\label{eq:n}
\end{equation} 
where $\mu_{D}$ is the cross-field diffusion coefficient and $\mu_{\nu}$ is the kinematic viscosity. The term $\Gamma$, defined as $\Gamma \equiv - \partial_{x} \ln(n_{0})$, measures the plasma density gradient, which is assumed constant, implying $n_{0} \propto \exp (-\Gamma x)$. The Poisson bracket is defined as: $[A, B]=\frac{\partial A}{\partial x} \frac{\partial B}{\partial y}-\frac{\partial A}{\partial y} \frac{\partial B}{\partial x}$. In these equations, the electrostatic potential $\phi$ is the stream function for the $\bm{E} \times \bm{B}$ velocity, represented by $\bm{u}=\nabla^{\perp} \phi$, where $\nabla^{\perp}=\left(-\frac{\partial}{\partial y}, \frac{\partial}{\partial x}\right)$. Thus, we have $u_x=-\frac{\partial \phi}{\partial y}$ and $u_y=\frac{\partial \phi}{\partial x}$, and the vorticity is given by $\omega=\nabla^2 \phi$.

The  variables are normalized as follows,  
$$
x/\rho_{\mathrm{s}} \rightarrow  x,\quad  \omega_{\mathrm{ci}} t \rightarrow t,  \quad e \phi / T_{\mathrm{e}} \rightarrow \phi,\quad n_{1} / n_{0} \rightarrow n \, ,
$$   where $\rho_s$ is the ion Larmor radius at electron temperature $T_{\mathrm{e}}$. It is  defined as $\rho_s = \frac{\sqrt{T_e/m_i}}{\omega_{ci}}$, where $\omega_{ci}$ is the ion cyclotron frequency.  Here $n_{1}$ and $n_{0}$ represent plasma density fluctuation and equilibrium density, while $\phi$ indicates electrostatic potential fluctuation.

The adiabaticity parameter $c$ measures the parallel electron response and is defined as,
\begin{equation}
c=\frac{T_{e} k_{\parallel}^{2}}{\mathrm{e}^{2} n_{0} \eta \omega_{c i}},
\end{equation}
where $\eta$ is the electron resistivity and $k_{\parallel}$ is the effective parallel wavenumber. The parameter $c$ determines the phase difference between the electrostatic potential and plasma density fluctuations.  For $c \gg 1$ (adiabatic limit), the model reduces to the Hasegawa--Mima equation, where electrons follow a Boltzmann distribution. In the hydrodynamic limit ($c \ll 1$), the system resembles the two-dimensional Navier--Stokes equation, where density fluctuations are passively influenced by the $\bm{E} \times \bm{B}$ flow. In the quasi-adiabatic regime ($c \sim 1$), there is a phase shift between potentials and densities, enabling particle transport.  In this paper, we focus on the quasi-adiabatic regime ($c = 0.7$), which is relevant to the edge plasma of tokamaks~\cite{Bos2010}. For more details about the simulation, we refer to \cite{Kadoch2022, Lin2024a, Lin2025}.

\subsection{Heavy impurity particle model, drag force, relaxation time}
Earlier research ~\cite{Futatani2008b, Futatani2008a, Futatani2009a} treated impurity particles as ``passive tracers'', assuming no inertia, meaning they closely followed the fluid flow. However, for heavier particles, this assumption might not hold due to their significant inertia, leading to noticeable inertial effects.

Impurity particles experience both the Lorentz force from electric and magnetic fields and a drag force due to momentum transfer from plasma ions. Positive plasma ions transfer momentum to heavy charged impurities through interaction with the electrostatic potential around the impurity particle, while electron momentum transfer is negligible due to their negligible mass. The effective cross-section for a single plasma ion colliding with a single impurity particle is given by \cite{MITOCW2003, Helander2005},
\begin{equation}
\sigma(v_r) =  \frac{m_i + m_p}{m_i^2 m_p}
\frac{Z^2e^4}{4 \pi \epsilon_0^2} \frac{  \ln \Lambda}{v_r^4} \, ,
\end{equation}
where  $m_i$ and $m_p$ represent the masses of plasma ion and the impurity particle respectively, and $Z$ is the charge number of the impurity particle.   Here  $v_r$ = $|\mathbf{v}_r|$ =  $|\mathbf{v} - \mathbf{v_p}|$ is the relative velocity between a plasma ion $\mathbf{v}$ and an impurity particle $\mathbf{v_p}$.  Moreover  $\ln \Lambda$ is the Coulomb logarithm. For detailed derivation, we refer to  \ref{appendix: derivation of cross section}.

To evaluate the drag force exerted by plasma ions on a single impurity particle, we consider the thermal (Maxwellian) distribution of plasma ions and the thermal velocity of impurity particles is negligible compared with plasma ions. Consequently, we adopt a drifting Maxwellian form for the plasma ion distribution,

\begin{equation}
f_i(\mathbf{v}) = n_i \left(\frac{m_i}{2 \pi T_i}\right)^{\frac{3}{2}} \exp \left[-\frac{m_i(\mathbf{v} - \mathbf{u})^2}{2 T_i}\right] \, ,
\end{equation}
where $\mathbf{u}$ is the drift velocity, $T_i$ is the plasma ion temperature. Given $\mathbf{v}_r = \mathbf{v}- \mathbf{v}_p \rightarrow \mathbf{v} = \mathbf{v}_r + \mathbf{v}_p $,  we can rewrite $f_i$  as,
\begin{eqnarray}
f_i(\mathbf{v}_r) &=& n_i \left( \frac{m_i}{2 \pi T_i} \right)^{\frac{3}{2}} \exp \left[-\frac{m_i \left( \mathbf{v}_r + \mathbf{v}_p - \mathbf{u} \right)^2}{2 T_i}\right] \nonumber \\
&=& n_i \left( \frac{m_i}{2 \pi T_i} \right)^{\frac{3}{2}} \exp \left[-\frac{m_i \left( \mathbf{v}_r^2 + 2 \mathbf{v}_r \cdot  (\mathbf{v}_p - \mathbf{u}) + (\mathbf{v}_p - \mathbf{u})^2 \right)}{2 T_i}\right] \, .
\end{eqnarray}

To approximate this integral, we assume $\mathbf{v}_p - \mathbf{u}$ is small relative to the typical thermal velocity of the plasma ion, $v_t \equiv \sqrt{T_i / m_i}$, then we express $f_i$ in terms of $\hat{\mathbf{v}}_r \equiv \mathbf{v}_r / v_t$, expanded to first order in $(\mathbf{v}_p - \mathbf{u}) / v_t \equiv (\hat{\mathbf{v}}_p - \hat{\mathbf{u}})$.
Thus we get,
\begin{eqnarray}
f_i &\approx& \frac{n_i}{(2 \pi)^{\frac{3}{2}} v_t^3} \left(1 + \hat{\mathbf{v}}_r \cdot (\hat{\mathbf{v}}_p - \hat{\mathbf{u}})\right) \exp [{-\frac{\hat{v}_r^2} {2}}] \nonumber \\
&=& \left(1 + \hat{\mathbf{v}}_r \cdot (\hat{\mathbf{v}}_p - \hat{\mathbf{u}})\right) f_0 \, ,
\end{eqnarray}
where $f_0 = (n_i/2 \pi^{\frac{3}{2}} v_t^3) \exp [{-\hat{v}_r^2/2}] $ represents the unshifted Maxwellian distribution. The total rate of momentum loss from the drifting Maxwellian plasma ions to a single impurity particle is,
\begin{eqnarray}
\frac{\mathbf{d p}}{d t} &=& \int f_i(\mathbf{v}_r) \sigma(v_r)v_r m_i \mathbf{v}_r \, d^3 \mathbf{v}_r \nonumber \\
&=& \int (1 + \hat{\mathbf{v}}_r \cdot (\hat{\mathbf{v}}_p - \hat{\mathbf{u}})) f_0 \sigma(v_r)v_r m_i \mathbf{v}_r \, d^3 \mathbf{v}_r\nonumber\\
&=& \int  (\hat{\mathbf{v}}_r \cdot (\hat{\mathbf{v}}_p - \hat{\mathbf{u}})) f_0 \sigma(v_r)v_r m_i \mathbf{v}_r \, d^3 \mathbf{v}_r \, , \nonumber\\
\label{eq:dp_dt1}
\end{eqnarray}
where $\int f_0 \sigma(v_r)v_r m_i \mathbf{v}_r \, d^3 \mathbf{v}_r = 0 $ because the overall integrand is an odd vector function of $\mathbf{v}_r$.
As detailed in \ref{appendix: integral}, the result of the equation (\ref{eq:dp_dt1}) is,
\begin{equation}
    \frac{d \mathbf{p}}{dt} =n_i \frac{m_i + m_p}{m_i m_p}
\frac{Z^2e^4}{6 \sqrt{2} \pi^{3 / 2} \epsilon_0^2} \frac{  \ln \Lambda}{v_t^3}  (\mathbf{v}_p - \mathbf{u}) \, . \nonumber
\end{equation}

The resulting drag force $\mathbf{F}_{drag}$ on an impurity particle due to plasma ions is equivalent to the negative momentum loss rate of the plasma ions,
\begin{equation}
\mathbf{F}_{drag} = -\frac{d \mathbf{p}}{dt} = n_i \frac{m_i + m_p}{m_i m_p}\frac{Z^2e^4}{6 \sqrt{2} \pi^{3 / 2} \epsilon_0^2} \frac{  \ln \Lambda}{v_t^3} (\mathbf{u}-\mathbf{v}_p) \, .
\end{equation}
For an individual impurity particle, the equation of motion is,
\begin{equation}
m_p \frac{d \mathbf{v}_p}{d t} = \mathbf{F}_{drag} + \mathbf{F}_L \, ,
\end{equation}
where \(F_L\) is the Lorentz force. Dividing  both sides by $m_p$, we get,
\begin{equation}
\frac{d \mathbf{v}_p}{d t} = \frac{\mathbf{u} - \mathbf{v}_p}{\tau_p} + \frac{Z e}{m_p} (\mathbf{E} + \mathbf{v}_p \times \mathbf{B}) \, ,
\label{eq: equation for particle velocity}
\end{equation}
where the relaxation time $\tau_p$ is defined as,
\begin{eqnarray}
\tau_p&=& \frac{1}{n_i} \frac{m_i m_p^2}{m_i + m_p}
\frac{6 \sqrt{2} \pi^{3 / 2} \epsilon_0^2}{Z^2e^4} \frac{v_t^3} {  \ln \Lambda} \nonumber\\
 &=& \frac{1}{n_i} \frac{m_p^2}{(m_i + m_p)\sqrt{m_i}}
\frac{6 \sqrt{2} \pi^{3 / 2} \epsilon_0^2}{Z^2e^4} \frac{T_i^{\frac{3}{2}}} {  \ln \Lambda} \, .
\end{eqnarray}
When \(m_p \gg m_i\), this simplifies to,
\begin{equation}
\tau_p = \frac{1}{n_i} \frac{m_p}{\sqrt{m_i}}
\frac{6 \sqrt{2} \pi^{3 / 2} \epsilon_0^2}{Z^2e^4} \frac{T_i^{\frac{3}{2}}} {  \ln \Lambda} \, .
\end{equation}
 This drag force reduces the velocity difference, allowing impurities to relax to the plasma flow velocity. The relaxation time  of the impurity particle  $\tau_p$,  characterizes the time scale that  impurity particles adjust to the plasma flow due to the drag force. For equation (\ref{eq: equation for particle velocity}), utilizing the same normalization applied to the HW equation, we obtain,
\begin{equation}
\frac{d \bm{v}_{p}}{d t}=\frac{(\bm{u}_{p}-\bm{v}_{p})}{\tau_p}+ \ \alpha (-\nabla \phi(\bm{x}_{p})+\bm{v}_{p} \times \bm{b}) \, ,
\label{eq: equation of motion 2}
\end{equation}
where the normalized relaxation time is given by $\omega_{\mathrm{ci}} \tau_p \rightarrow \tau_p$ and $\alpha=Zm_i/m_p$, with $m_i$ denoting the ion mass of the plasma, and $\bm{b}$ being the unit vector aligning with the magnetic field's direction, whereas $\bm{u}_{p}$,  and $\nabla \phi(\bm{x}_{p})$ represent  the fluid's velocity and the potential's gradient at the position $\bm{x}_{p}$ of the impurity particle, respectively.

Together with the equation,
\begin{equation}
\frac{d \bm{x}_{p}}{dt} = \bm{v}_{p}  \, ,
\label{eq: equation of motion 3}
\end{equation}
the trajectory of the impurity particles can be determined.
In this study, we focus on tungsten as the impurity particle of interest. Tungsten is a heavy element that can originate from materials exposed to high heat fluxes in devices such as ASDEX-Upgrade~\cite{Krieger1999} and EAST~\cite{Yao2015},  expecting the same situation for ITER~\cite{Pitts2009}.  We investigate charge numbers $Z=3,10,20$ and 60 to represent low, medium, and highly charged states of tungsten ions. The parameters used to calculate the relaxation time are listed in table~\ref{tab:relaxation_time_parameters}. The normalized relaxation times and $\alpha$ for different values of $Z$ for tungsten ions are presented in table~\ref{tab:taup_alpha}. Table~\ref{tab:taup_alpha}  demonstrates that $1/\tau_p$ is indeed much larger than $\alpha$ for the parameters considered. Let us note that  from Eq.~\ref{eq: equation of motion 2}, when $1/\tau_p$ (flow drag) significantly exceeds $\alpha$ (Lorentz force), particle dynamics are primarily controlled by fluid drag. 

\begin{table}
\caption{\label{tab:relaxation_time_parameters}Parameters for calculating relaxation time include: $B$ as the magnetic field, $n_i$ representing the plasma ion density, $\ln \Lambda$ for the Coulomb logarithm, and $T_i$ as the plasma ion temperature.}
\begin{indented}
\lineup
\item[]\begin{tabular}{@{}llll}
\br
$B(T)$ & $n_i(m^{-3})$ & $\ln \Lambda$ & $T_i(eV) $ \\ 
\mr
1   & $1 \times 10^{19}$  & 10 & 0.1 \\
\br
\end{tabular}
\end{indented}
\end{table}

\begin{table}
\caption{\label{tab:taup_alpha} Normalized relaxation time $\tau_p$, $\alpha$ and  Stokes number $St$ for different charge numbers $Z$ of tungsten ions.}
\begin{indented}
\lineup
\item[]\begin{tabular}{@{}lcccccc}
\br
$Z$  & 3 & 10 & 20 & 60 \\
\mr
$\tau_p$  & 3.20  & 0.29 & 0.07 & 0.01 \\
\mr
$\alpha$ & 0.03 & 0.11 & 0.22 & 0.66 \\
\mr
$St$ & 9.14 & 0.83 & 0.20 & 0.03 \\
\br
\end{tabular}
\end{indented}
\end{table}

\subsection{Stokes number}
The particle distribution will be influenced by the relative relation between the relaxation time $\tau_p$ of the particle  and the characteristic time scale of turbulence $\tau_{\eta}$, which indicates how fast the fluid flow changes. In the fluid mechanics community, the Stokes number $St$ \cite{Brennen2005, Israel1982} is widely used to quantify particle inertia in fluid flow. The Stokes number $St$ is defined as,
\begin{equation}
St=\frac{\tau_p}{\tau_\eta}  \, .  
\end{equation}
A high Stokes number means that a particle's movement is primarily driven by inertia, causing it to keep its own direction despite fluid changes. A small Stokes number means that the particle’s movement is more strongly influenced by the surrounding fluid, causing it to follow the fluid's flow closely and adapt quickly to any changes in the fluid velocity.  In this  work, the characteristic time scale of turbulence, $\tau_{\eta}$, is defined as $1 / \sqrt{2 Z_{\rm m s}}$, with $Z_{\rm m s}$ denoting
one half of the mean-square vorticity.
In the simulation we have $\tau_{\eta} = 0.35$.   The Stokes numbers $St$ for different charge numbers $Z$ of tungsten ions are listed in table~\ref{tab:taup_alpha}.

\subsection{Multiscale geometrical Lagrangian statistics:  Curvature and scale-dependent curvature angle}
In the context of Lagrangian dynamics, the evolution over time of the impurity particle's position is governed by equations (\ref{eq: equation of motion 2}) and (\ref{eq: equation of motion 3}), given the initial position, $\bm{x}_p(t=0)$ $=\bm{x}_{p0}$. Particle movement over time is simulated using a second-order Runge--Kutta method and linear interpolation is used for computing the fluid velocity at the particle position. 

The curvature $\kappa$ is defined as,

\begin{equation}
\label{eq: kappa}
    \kappa=\frac{a_n}{\left\|\bm{v}_p\right\|^2} \, ,
\end{equation}
where $a_n$ represents the normal component of the Lagrangian acceleration that is perpendicular to the velocity vector. This component achieves zero when the velocity and acceleration vectors align. We employed the same geometrical Lagrangian analysis as utilized in Navier--Stokes turbulence \cite{Kadoch2017} to investigate the heavy impurities behavior in  drift-wave turbulence.

To understand the  multiscale dynamics of heavy impurities in drift-wave turbulence from a Lagrangian framework, we statistically evaluate directional motion within stochastic trajectories at different time intervals using the curvature angle \cite{Bos2015, Burov2013}. We introduce
time increments of the Lagrangian particle position,
\begin{equation}
 \delta \bm{x}_p\left( t, \tau\right)=\bm{x}_p(t)-\bm{x}_p(t-\tau) \, ,
 \end{equation}
\begin{figure}
\centering
\includegraphics[width=0.8\textwidth]{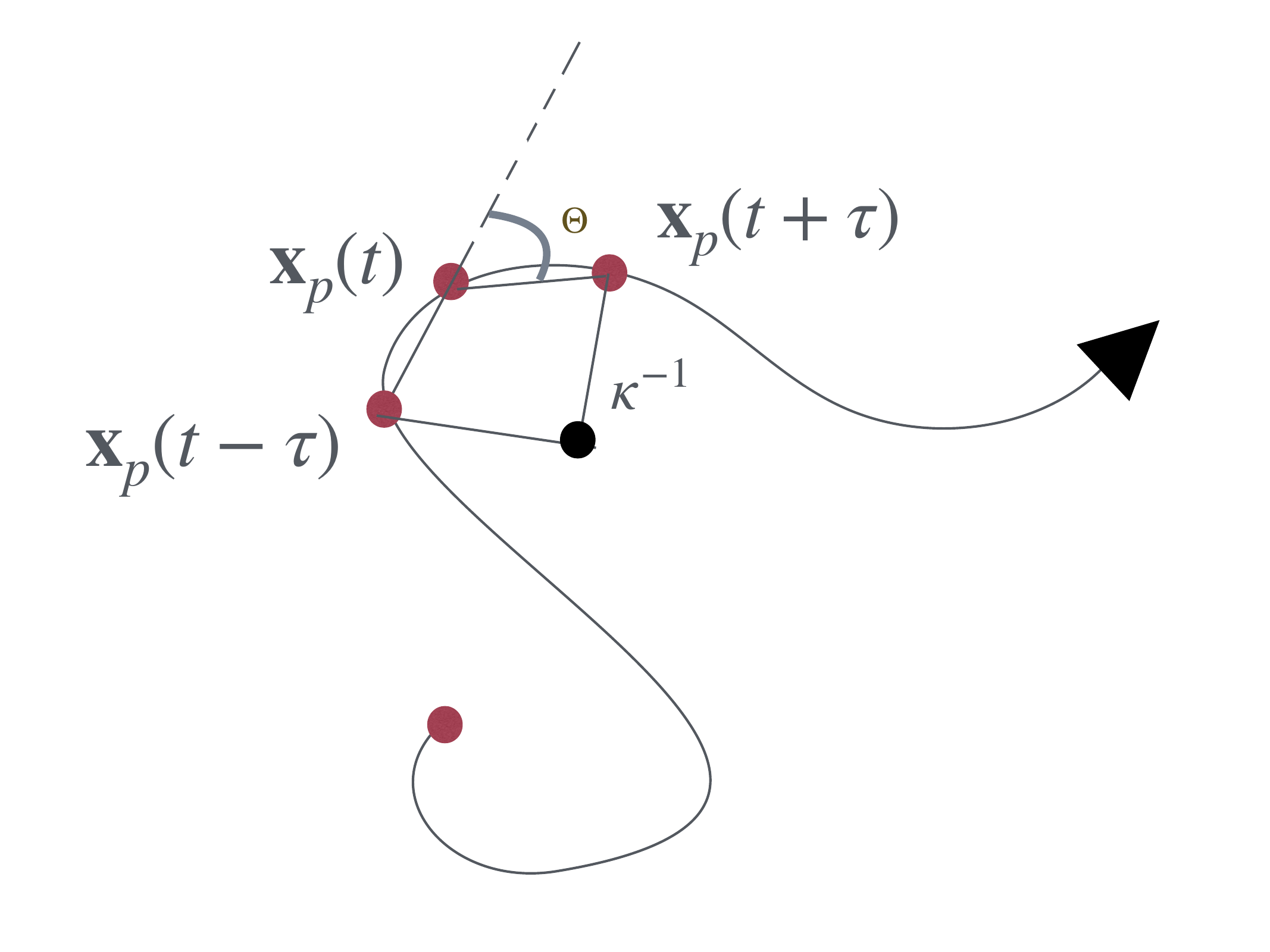}
\caption{Illustration of the curvature angle $\Theta$ and  the curvature $\kappa$.   $\bm{x}_p(t-\tau)$,  $\bm{x}_p(t)$ and  $\bm{x}_p(t+\tau)$ marked by red dots are the positions of the impurity particle $p$ at time instants $t - \tau$,  $t$, and $t + \tau$, respectively. The inverse curvature $\kappa^{-1}$ corresponds to the radius of the unique circle passing through these three points, providing a discrete approximation at scale $\tau$ of the local trajectory curvature.}
\label{fig: Theta}
\end{figure}
where $\bm{x}_p(t)$is the impurity particle location at time $t$. The cosine of the angle $\Theta(t, \tau)$ between consecutive particle increments, originally introduced in \cite{Burov2013} for random walks and complex biological systems and examined in both three-dimensional homogeneous isotropic turbulence \cite{Bos2015} and two-dimensional homogeneous isotropic and confined turbulence \cite{Kadoch2017}, is given by, 
\begin{equation}\label{eq:cos_theta} \cos (\Theta (t, \tau))=\frac{\delta \bm{x}_p(t, \tau) \cdot \delta \bm{x}_p( t+\tau, \tau)}{|\delta \bm{x}_p(t, \tau)||\delta \bm{x}_p( t+\tau, \tau)|} \, .
\end{equation}

This quantity describes the directional change of a particle for a specific scale, \textit{i.e.}, the time increment size $\tau$, allowing for multiscale geometric analysis. Figure \ref{fig: Theta} shows an illustration of the scale-dependent curvature angle $\Theta$. The ensemble and time average of the absolute curvature angle is denoted by,
\begin{equation}
    \theta(\tau) \equiv\langle | \Theta(t, \tau)| \rangle.
\end{equation}

The curvature angle $\Theta$ is associated with the curvature $\kappa$ as specified in equation (\ref{eq: kappa}). By normalizing the scale-dependent curvature angle $\Theta$, which is done by dividing $\Theta(t, \tau)$ by $2 \tau\|\bm{v_p}\|$, we obtain the finite-time curvature,
\begin{equation}
    K(t, \tau) = \Theta(\tau) /(2 \tau\|\bm{v_p}\|) \, .
    \label{eq: K}
\end{equation}
The curvature $\kappa$ is obtained by taking the limit as the time increment $\tau$ approaches zero.

\section{Results}\label{sec:results}

We performed DNS computations using a fully dealiased pseudo-spectral method  at resolution $N_x \times N_y = 1024^2$ grid points in a double periodic domain of size length $  64\rho_s $ with time step $\Delta t = 5 \times 10^{-4}$. Diffusivity and kinematic viscosity were chosen equally, $\mu_\nu = \mu_D = 5 \times 10^{-3}$, the mean plasma density gradient $\Gamma$ was set to 1 and the adiabaticity parameter $c=0.7$. When the flow became statistically stationary, we injected $10^6$ randomly distributed point particles, considering four Stokes numbers, cf. table~\ref{tab:taup_alpha}, 
and likewise fluid particles ($St=0$) as reference. 
Computations were run for about four hundred eddy turn over times to acquire sufficient statistics.
Details on the numerical simulation for fluid particles can be found in \cite{Kadoch2022}.  

\begin{figure}
\centering
\includegraphics[width=1.0\textwidth]{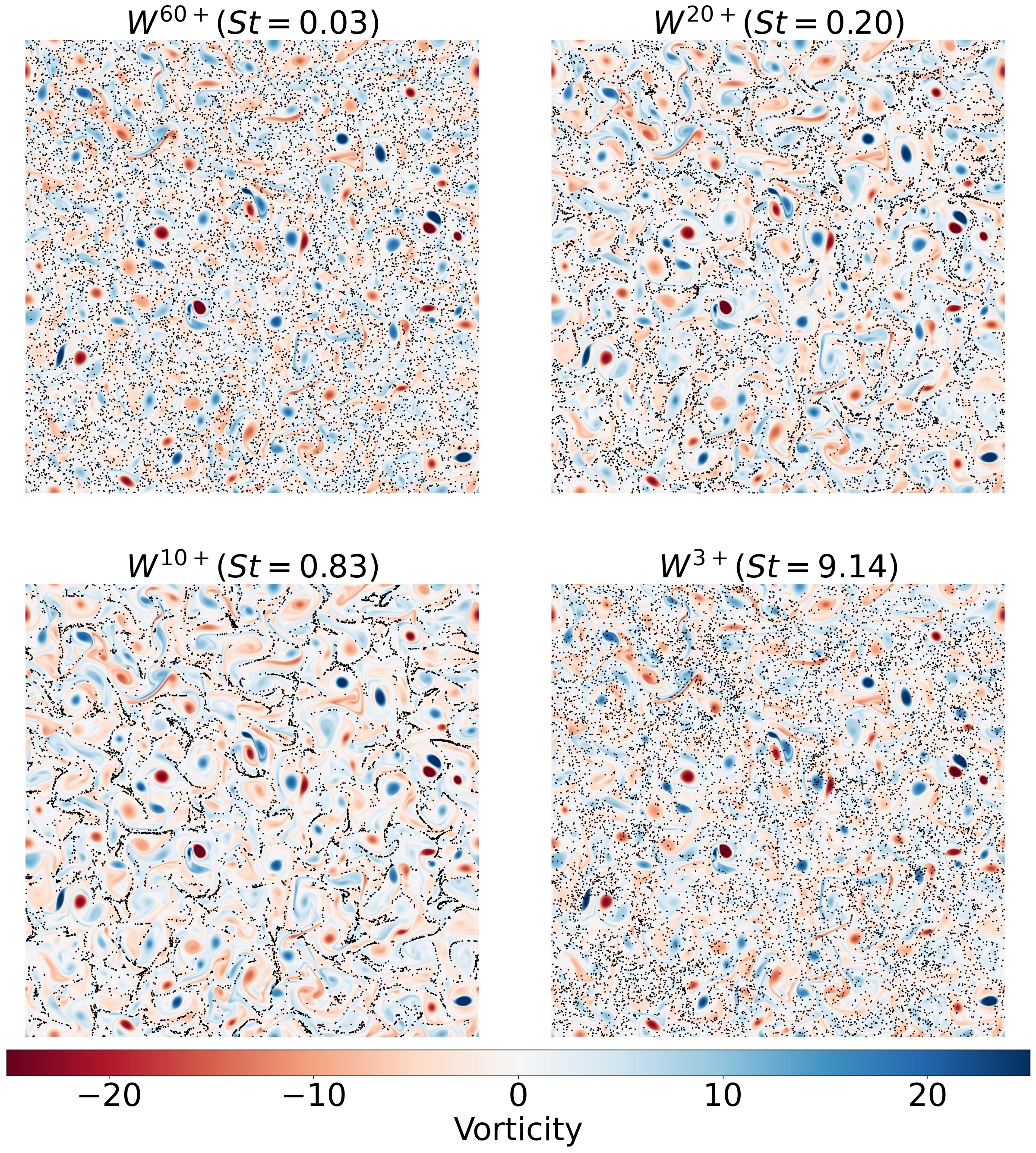}
\caption{Vorticity fields with $10^4$ impurity particles (sampled from a total of $10^6$) superimposed for $W^{60+} (St = 0.03)$, $W^{20+} (St = 0.20)$, $W^{10+} (St = 0.83)$, and $W^{3+} (St = 9.14)$ in the statistically steady state.}

\label{fig: vorticity and particles}
\end{figure}

Figure \ref{fig: vorticity and particles} shows the  vorticity fields with $10^4$ impurity particles (sampled from a total of $10^6$) superimposed for $W^{60+} (St = 0.03)$, $W^{20+} (St = 0.20)$, $W^{10+} (St = 0.83)$, and $W^{3+} (St = 9.14)$ in the statistically steady state. From figure \ref{fig: vorticity and particles}, clustering of impurity particles is evident for $W^{20+} (St = 0.20)$ and $W^{10+} (St = 0.83)$. For $W^{60+} (St = 0.03)$ and $W^{3+} (St = 9.14)$, the degree of inhomogeneous particle distribution is smaller.  In our numerical simulations, we use a globally fixed Stokes number $St$ for each type of impurity rather than a locally varying one. The maximum vorticity magnitude in the system is $|\omega|_{\max} = 23$, which corresponds to a minimum local time scale of  $ \tau_{\eta, \min} = \frac{1}{|\omega|_{\max}}. $  This results in :  $ |\omega|_{\max} \tau_\eta \approx 8. $ The  variations in vorticity enhance inertial effects in high-vorticity regions. At low Stokes numbers, the particle relaxation time is negligible compared to the eddy turnover time, \textit{i.e.}, $\tau_p \ll \tau_\eta(St \sim 0)$. Their relaxation time is sufficiently short, allowing them to rapidly align their velocity with flow changes. 
When the particle relaxation time is on the order of the eddy turnover time $\tau_p \sim \tau_\eta (S t \sim 1)$, particles experience the vortex-induced centrifugal effect.   Vortices can accelerate particles enough to be ejected from the vortices due to inertia, leading to movement toward regions with lower vorticity and resulting in clear particle clustering. Let us note that in regions of impurity clustering, there is a risk of violating the quasi-neutrality assumption fundamental to the HW model~\cite{Naulin2006}. This assumption requires that impurity density remains smaller than the bulk current divergence on the characteristic drift-wave timescale, which may not hold in areas of intense clustering. Nevertheless, as highlighted in~\cite{Naulin2006}, treating impurities as test particles is still valuable for understanding impurity transport tendencies within a given flow field. The clustering effects observed in our simulations offer insights into impurity dynamics.

\begin{figure}
 \centering
\includegraphics[width=0.8\textwidth]{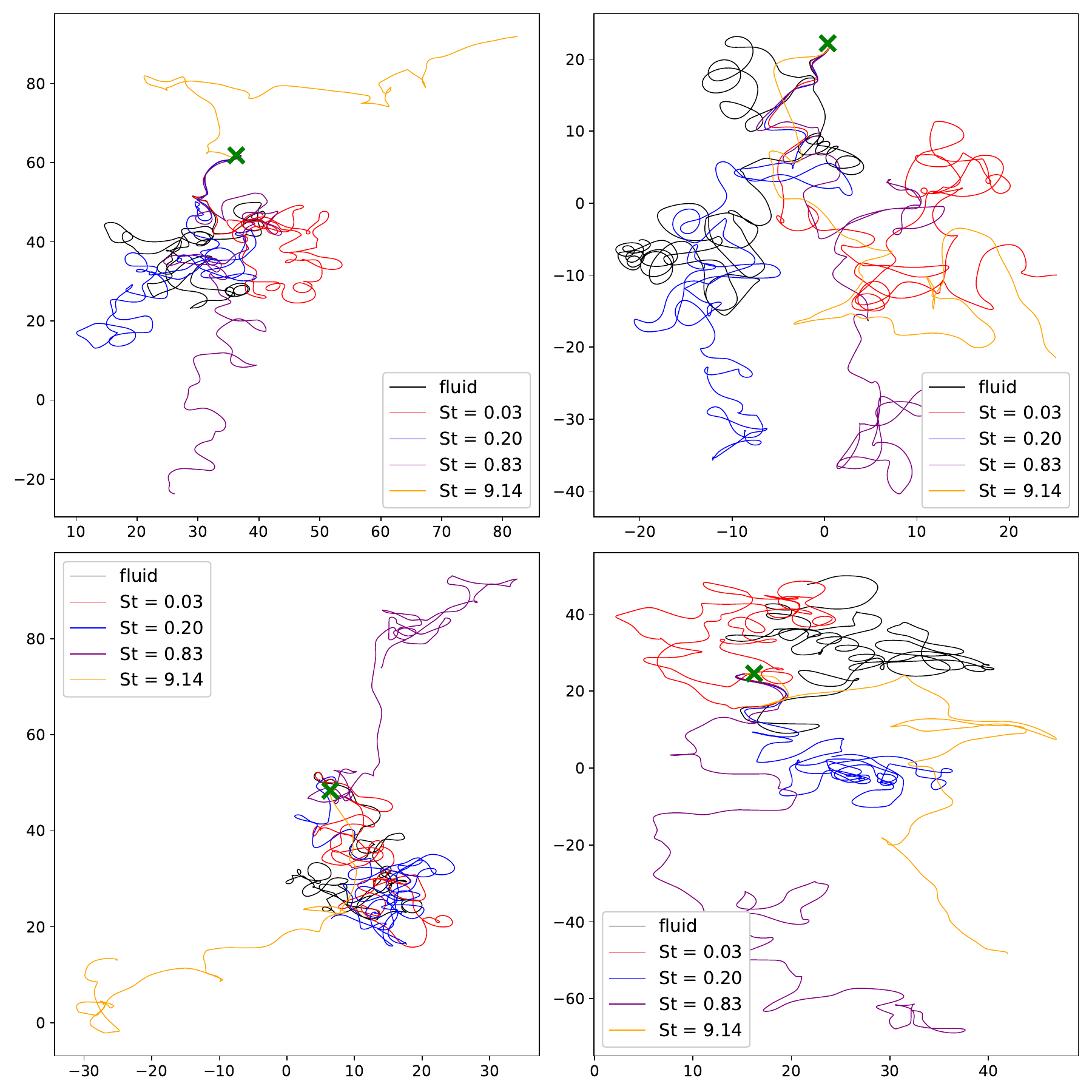}
\caption{Four plots are randomly selected for showing the trajectories of particles with different Stokes numbers, with their common initial position marked by a green cross. The periodic domain $[0, 64] \times [0, 64]$ has been extended when necessary to ensure continuity in trajectory visualization. For clarity, only the relevant portion of the domain is displayed.}
\label{fig: trajectory}
\end{figure}

 \begin{figure}
    \centering
    {\includegraphics[width=0.48\textwidth]{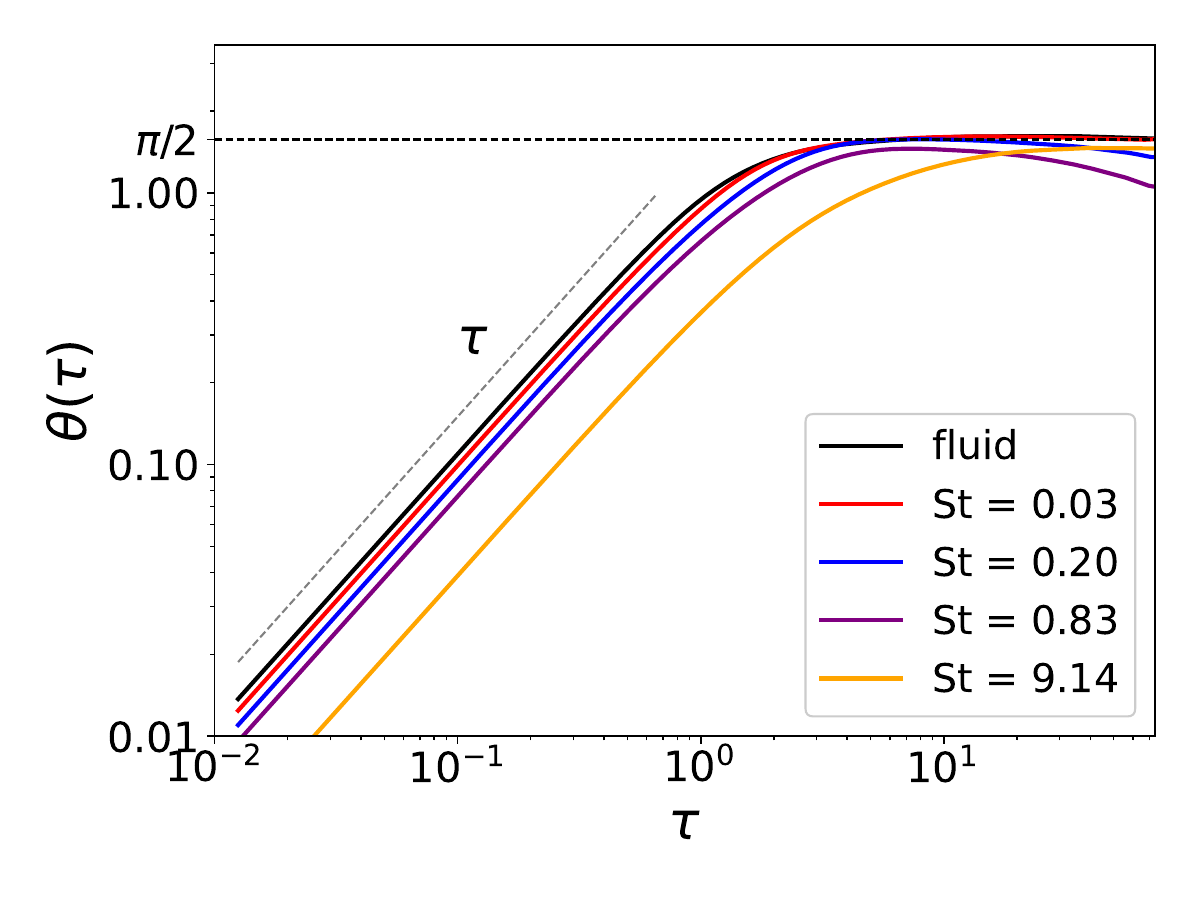}\quad
    \includegraphics[width=0.48\textwidth]{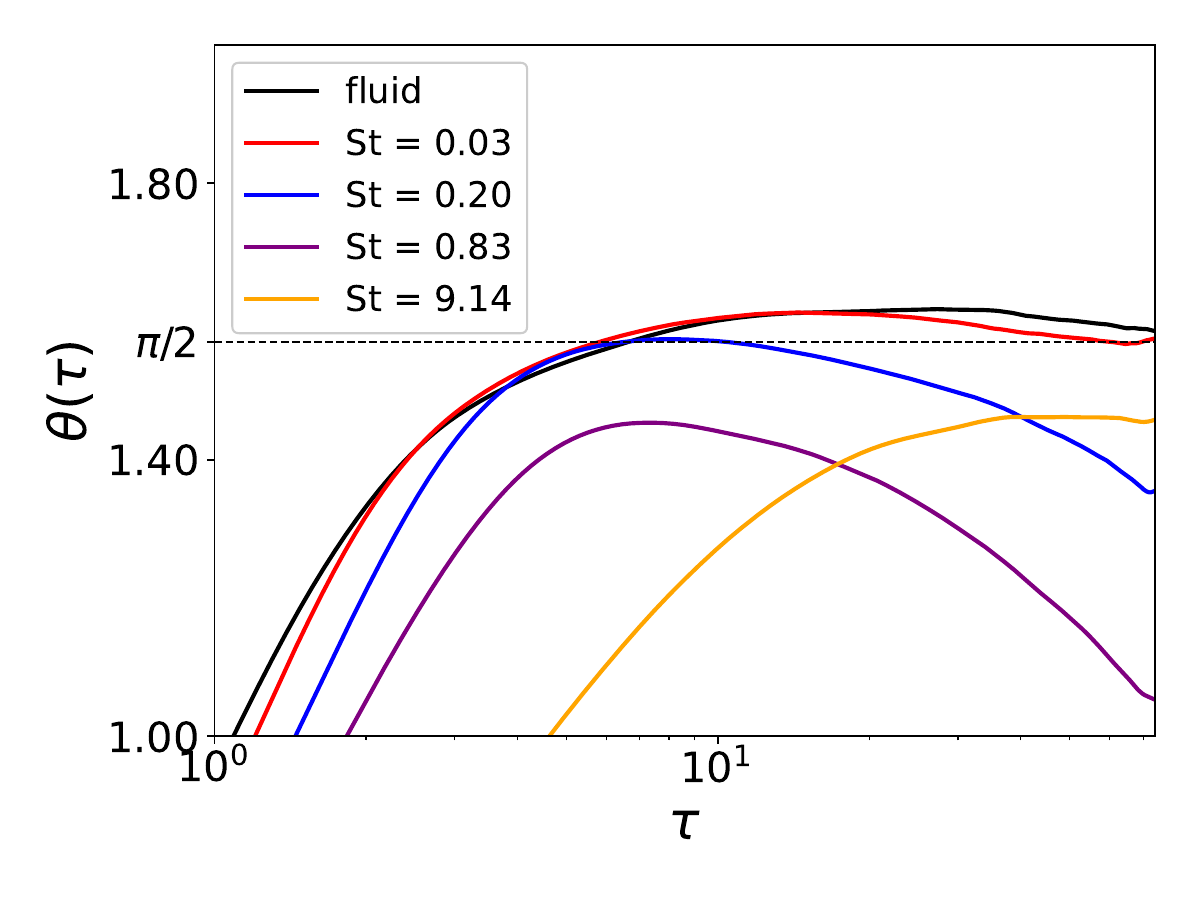} }
    \caption{Left: Mean curvature angle $\theta(\tau)$ versus $\tau$ for different Stokes numbers $St$. Right: A magnified view for large $\tau$. }    
 \label{fig: mean angle}
\end{figure}

In Figure \ref{fig: trajectory}, four plots are randomly selected for showing the trajectories of particles with different Stokes numbers, with their common initial position marked by a green cross.  For visualization clarity, the periodic computational domain  has been extended by adding one (or several) period(s)  to present  continuous  paths  of the particles.The same periodic prolongation has been applied for computing the curvature statistics.  
 
For fluid tracers ($St=0$) and particles with a small Stokes number ($St=0.03$), the trajectories exhibit random motion, as illustrated in figure \ref{fig: trajectory}. As the Stokes number increases,  changes in particle behavior become evident. Especially at $St=0.83$, the trajectories begin to show characteristics of ``Lévy flight'',  where particles are intermittently trapped in regions of the flow before abruptly jumping to new locations. “Levy-flight” trajectories characteristics have important implications for particle dispersion. For the particles with a high Stokes number ($St=9.14$), inertia dominates their dynamics. The particles are less influenced by the fluid flow and tend to maintain their momentum.

Figure~\ref{fig: mean angle} illustrates the mean curvature angle versus the time increment size $\tau$ for different Stokes numbers in a double logarithmic scale. For short $\tau$, a linear scaling is observed:  $\theta(\tau) \equiv\langle | \Theta(t, \tau) |\rangle = \langle 2\|\mathbf{v}_p\| \kappa \rangle \tau$ as $\tau \rightarrow 0$, aligning with the limit expressed in equation (\ref{eq: K}). This linear scaling behavior is similar to the findings in hydrodynamics \cite{Kadoch2017}. For small time increment size $\tau$, the mean curvature angle $\theta(\tau)$ is expected to approach zero, indicating that particles are moving straight. For larger time increment size $\tau$,  the mean curvature angles  $\theta(\tau)$  of fluid tracers ($St = 0$) and impurity particles with low Stokes number ($St = 0.03$), approach an asymptotic value of $\pi/2$, which corresponds to the mean angle of three random points and their  decorrelation \cite{Kadoch2017}. For particles with intermediate and high Stokes number ($St = 0.20, 0.83, 9.14$), they exhibit an asymptotic mean curvature angle less than $\pi/2$. This is because for particles with intermediate Stokes numbers ($St = 0.20, 0.83$), intermittent trapping and sudden jumps result in preferential movement in certain directions, leading to reduced overall directional changes over longer time scales. This effect is visible in Figure \ref{fig: trajectory}.  For high Stokes number particles, inertia causes them to follow smoother paths with minimal influence from flow fluctuations, further reducing curvature angles over time, as evidenced in  Figure \ref{fig: trajectory}.

Additionally, as shown in Figure~\ref{fig: mean angle}, the increase in Stokes numbers clearly shifts the curves to lower values, indicating reduced curvature angles for higher $St$ particles, thus reflecting their smoother paths  and a reduced tendency for U-turns.  Adding inertia can be seen as a low pass filter. The high Stokes number causes the particles to ``smooth out'' the effects of vortices and high-frequency variations

\begin{figure}
\centering
\includegraphics[width=0.8\textwidth]{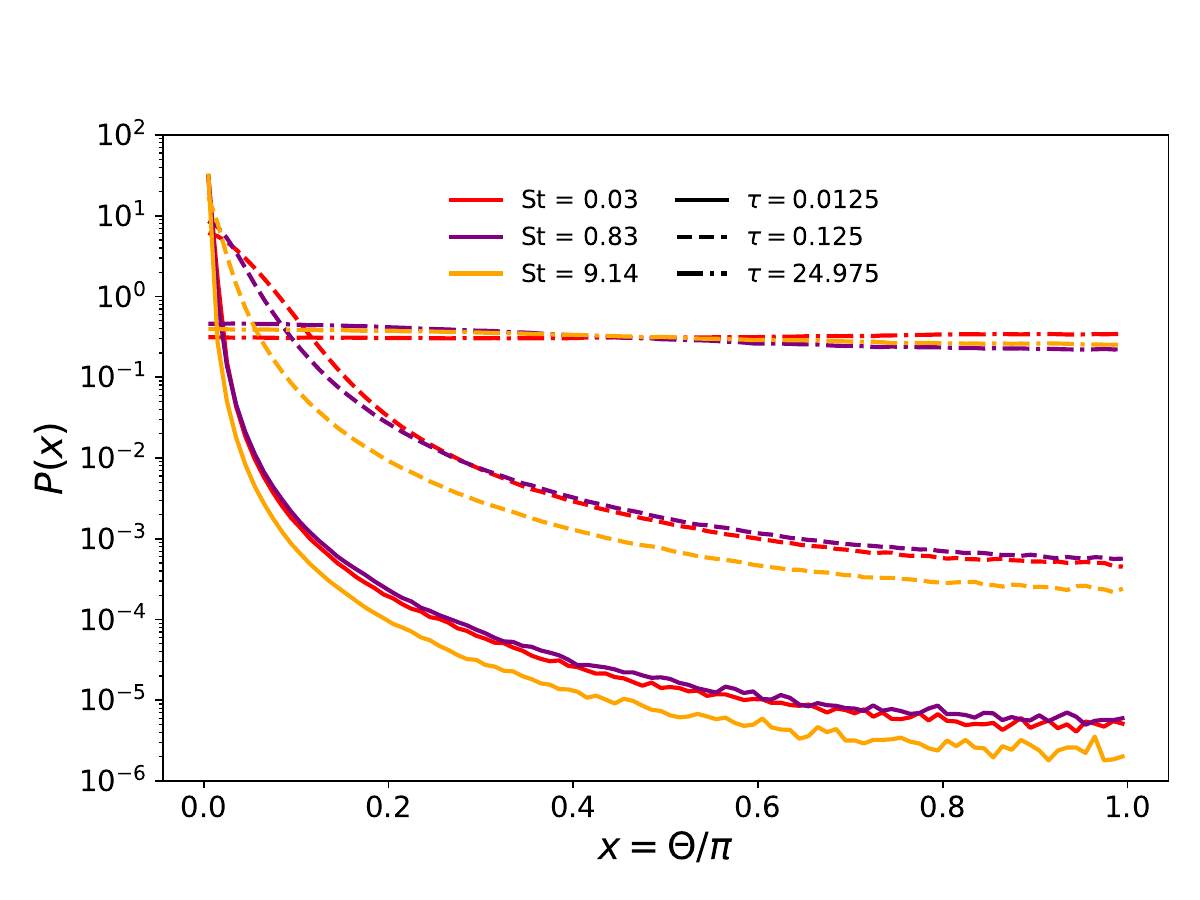} 
\caption{PDF of the normalized curvature angle $x = \Theta/\pi$ for different Stokes numbers $St$ and different time lags $\tau$.  
}
\label{fig: PDF of the of the curvature angle}
\end{figure}

\begin{figure}
\centering
\includegraphics[width=0.8\textwidth]{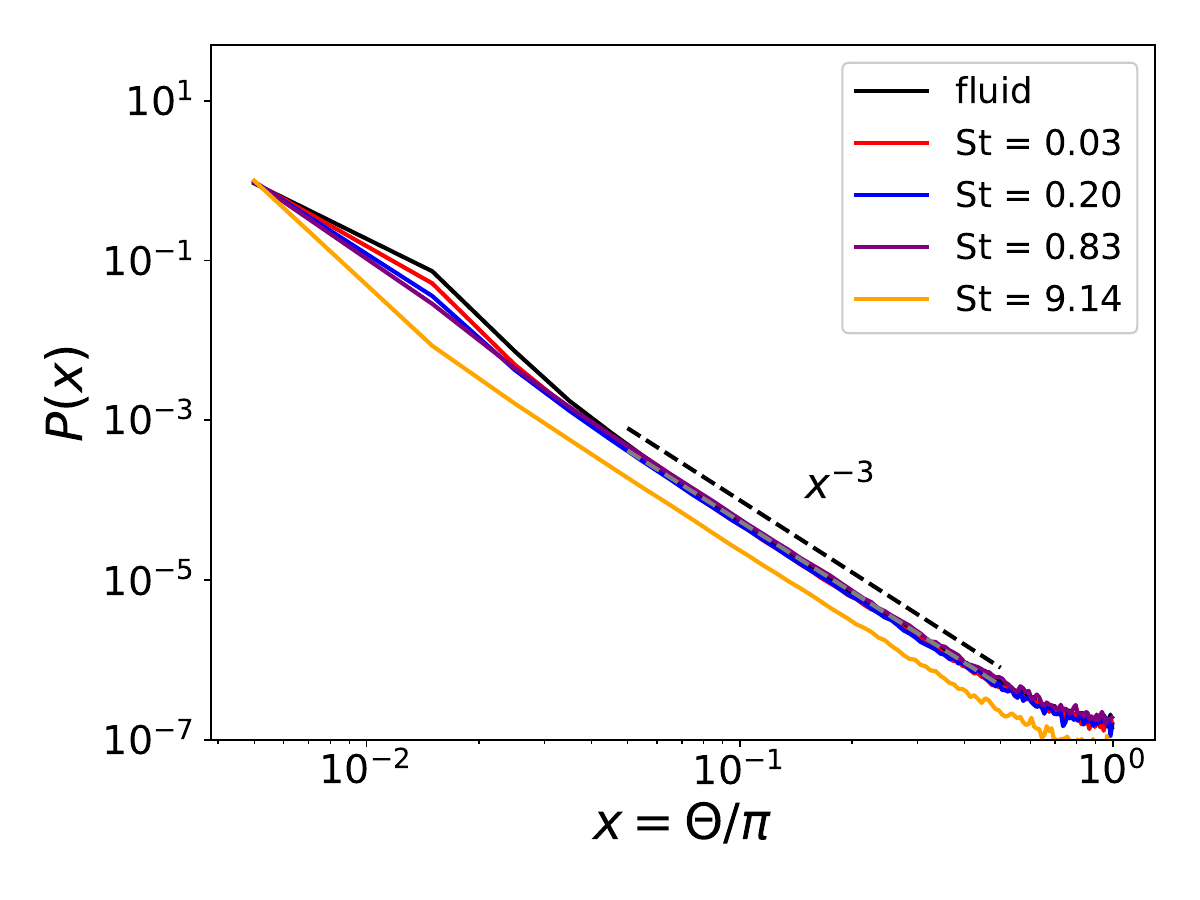} 
\caption{PDF of the normalized curvature angle $x = \Theta/\pi$ for different Stokes numbers $St$ and for  time lags $\tau =0.0125$.}
\label{fig: PDF of the curvature angle for tau =0.0125}
\end{figure}

Figure~\ref{fig: PDF of the of the curvature angle} presents the probability density function (PDF)  of the normalized curvature  angle $\Theta/\pi$ in a log-lin representation, focusing on three Stokes numbers $St = 0.03, 0.83$ and 9.14, across varying time increment size $\tau$. For all the three Stokes numbers, as the time increments increase, the PDF transitions from a highly peaked, fast decaying  distribution to a uniform distribution.  For $S t=0.83$ and $S t = 9.14$, the PDFs at large time increments approach a nearly uniform distribution but with an inclination, indicating a preference for smaller angles for long time increments. This inclination corresponds to the asymptotic mean curvature angle less than $\pi / 2$, as seen in figure \ref{fig: mean angle}.  For shorter increments, as expected, a peak at $\Theta=0$ indicates a high likelihood of particles moving straight.
For $St = 9.14$, the PDF remains sharper for longer time increments, indicating that high Stokes number particles preserve their initial motion direction over time. This behavior reflects their inability to rapidly adjust to changes in the turbulent fluid velocity due to their high relaxation time. Figure \ref{fig: PDF of the curvature angle for tau =0.0125} shows the PDFs of the normalized curvature angle $ \Theta/\pi$ for different Stokes numbers $St$ and for time lag $\tau =0.0125$.   The PDFs exhibit a power-law scaling with a slope close to -3 at large  $ \Theta/\pi$.

\section{Conclusions} \label{sec:conclusion}
We developed multiscale diagnostics to analyze the Lagrangian statistics of heavy impurities, specifically tungsten, in drift-wave turbulence using the HW equations. Employing classical Coulomb scattering theory, we derived the relaxation times of heavy impurities and found that particles with lower charge numbers have higher relaxation times, leading to higher Stokes numbers in the turbulent flow. Let us note that in fluid mechanics, heavier particles generally correspond to higher Stokes numbers. In our research on fusion plasma, we focus on tungsten, where the particle mass is fixed, and the Stokes number is determined by the charge number.Our study revealed that the Stokes number significantly influences the spatial impurity
distribution within drift-wave turbulence:

\begin{itemize}
    \item Low Stokes numbers (high charge states): Particles are closely following the fluid flow and exhibit certain degree of inhomogeneous 
distribution. Their quick velocity relaxation allows them to adjust rapidly to changes in the flow.
    \item Intermediate Stokes numbers (intermediate charge states): Particles exhibit clear clustering.   Their relaxation time is comparable to the eddy turnover time, causing centrifugal effects that eject particles from vortices and lead to accumulation in regions of low vorticity.
    \item High Stokes numbers (low charge states): Particles move ballistically and are less influenced by the turbulent flow, leading to a nearly  random distribution without significant clustering.
\end{itemize}

By applying multiscale curvature angle analysis to particle trajectories of impurities, we observed:
\begin{itemize}
    \item Small time lags: The mean curvature angle scales linearly, indicating that particles move in nearly straight paths over short timescales. 
    \item Large time lags: Low Stokes number particles approach an asymptotic mean curvature angle of $\pi/2$, reflecting trajectory decorrelation due to turbulence. High Stokes number particles  maintain mean curvature angles less than $\pi/2$, which means they do not visit all directions with the same probability, there is a mean direction.
\end{itemize}

The PDFs of curvature angles reveal that increasing the Stokes number delays the transition from small to large-scale statistical behavior. For higher Stokes numbers, the PDFs remain peaked at small curvature angles for longer time increments, indicating that these particles retain their initial motion direction over time. This reflects their slower response to turbulent fluid changes due to their higher relaxation time. Low-Z heavy impurities (corresponding to high Stokes numbers) tend to spread more rapidly and may possibly penetrate deeper into the plasma. Their reduced sensitivity to plasma flow might allow them to move more freely toward the plasma core, where their accumulation could degrade confinement by increasing radiative losses. Therefore, future work should examine the intermittency and diffusion properties of heavy impurities in turbulent plasmas to better understand these potential effects. Based on the observed trajectories of particles with varying Stokes numbers, further analyses can be performed to understand the underlying dynamics and statistical properties of their motion. For example, exit time and return time analyses can reveal how inertia affects the likelihood of particles escaping or returning to specific regions of the flow, providing insights into the intermittent trapping and sudden jumps observed in \cite{Benkadda1994, Benkadda1997}. 

\section*{Data availability statement}

The data that support the findings of this study are available upon reasonable request from the authors.  

\section*{Conflict of Interest}

The authors have no conflicts to disclose. 

\section*{Acknowledgements}

ZL and KS  acknowledge financial support from I2M. ZL, KS and SB acknowledge the financial support from  the French Federation for Magnetic Fusion Studies (FR-FCM) and the Eurofusion consortium,  funded by the  Euratom  Research and Training Programme under Grant Agreement No. 633053. The views and opinions expressed herein do not necessarily reflect those of the European Commission. ZL, BK and KS acknowledge partial funding from the Agence Nationale de la Recherche (ANR), project CM2E, grant ANR-20-CE46-0010-01. Centre de Calcul Intensif d’Aix-Marseille is acknowledged for providing access to its high performance computing resources.

\section*{ORCID iDs}

\noindent
Zetao Lin \url{ https://orcid.org/0009-0001-3033-5106}\\
Benjamin Kadoch \url{ https://orcid.org/0000-0001-9346-1399}\\
Sadruddin Benkadda \url{https://orcid.org/0000-0002-0717-9125}\\
Kai Schneider \url{https://orcid.org/0000-0003-1243-6621}
\appendix
\section{Derivation of effective cross section}\label{appendix: derivation of cross section}

\begin{figure}
\centering
\includegraphics[width=0.7\textwidth]{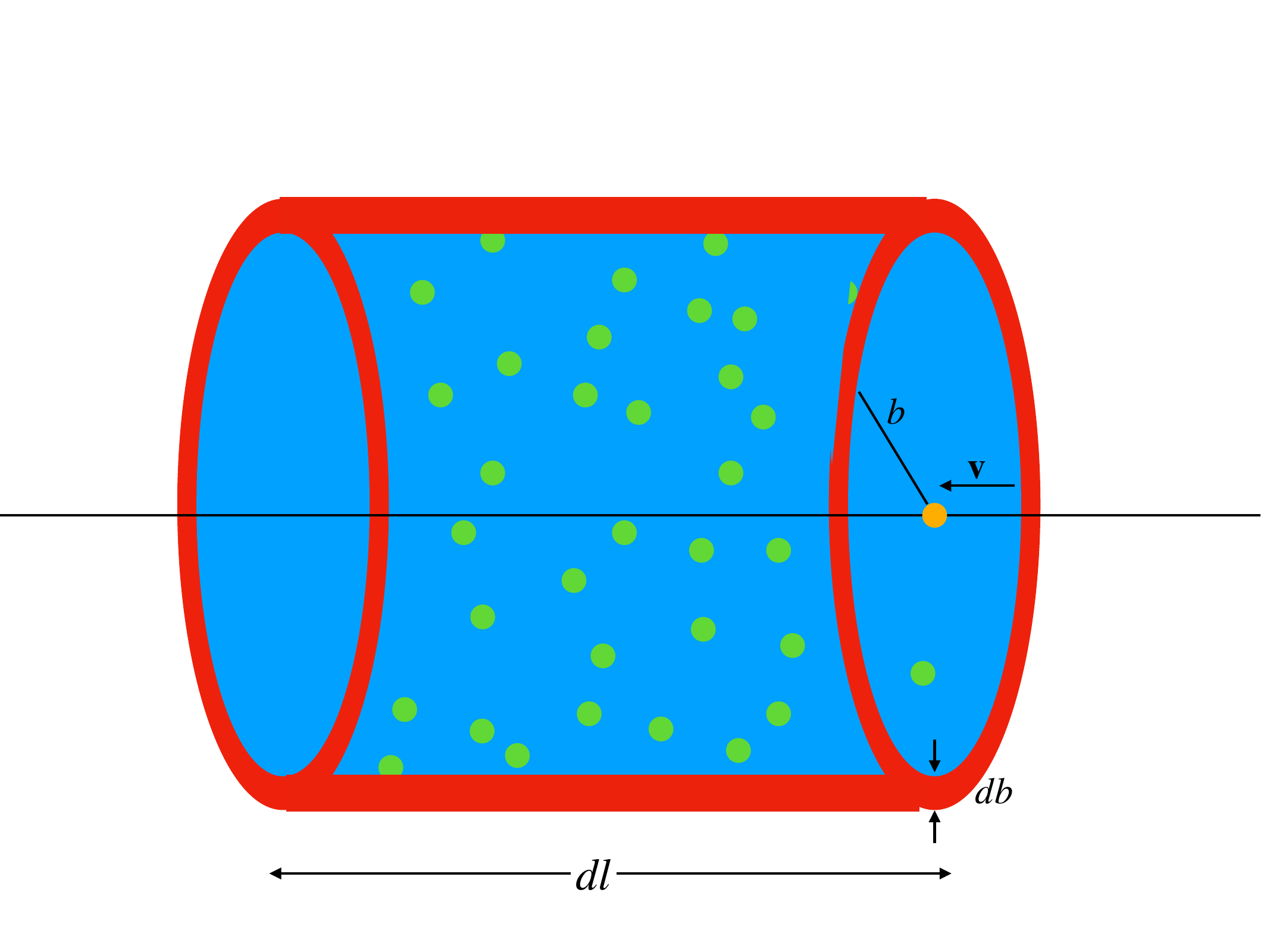}
\caption{Illustration of the annular volume with heavy impurities (green) and a moving plasma ion (orange).} 
\label{fig: annular volume}
\end{figure}

Consider a single plasma ion with mass $m_i$ and velocity $\mathbf{v}$, moving through a field of impurity targets with density $n_p$, where each impurity particle has charge $Ze$, mass $m_p$, and velocity $\mathbf{v_p}$. The plasma ion moves along an annular path with radius $b$ and thickness $db$ over an infinitesimal distance $d\ell$. This path corresponds to a volume element $d\ell \, 2 \pi b \, db$, within which the plasma ion encounters a total number of impurity targets given by $n_p d \ell 2 \pi b d b $. Figure \ref{fig: annular volume} illustrates the annular volume corresponding to $db$. Using  the classical Coulomb scattering theory, a collision with an impact parameter $b$, $\mathbf{v}$ changes by an amount \cite{Helander2005} as follows,

\begin{equation}
    \Delta v_x \simeq-\left(1+\frac{m_i}{m_p}\right)\left(\frac{Ze^2}{2 \pi \epsilon_0 m_i}\right)^2 \frac{1}{2 b^2 v_r^3} \, ,
\end{equation}
where $v_r$ = $|\mathbf{v}_r|$ =  $|\mathbf{v} - \mathbf{v_p}|$ is the relative velocity between a plasma ion and an impurity particle. The rate of momentum loss of the plasma ion along the path $d\ell$ is thus expressed by the integral,

\begin{eqnarray}
\frac{d p}{d \ell} & =  \int_{b_{min} }^{b_{max}} m_i  \Delta v_x n_p 2 \pi b d b \nonumber \\
& =   \int_{b_{min} }^{b_{max}} m_i \left(1+\frac{m_i}{m_p}\right)\left(\frac{Ze^2}{2 \pi \epsilon_0 m_i}\right)^2 \frac{1}{2 b^2 v_r^3}  n_p 2 \pi b d b \nonumber \\
& =   n_p \frac{m_i + m_p}{m_i m_p}
\frac{Z^2e^4}{4 \pi \epsilon_0^2} \frac{  \ln \Lambda}{v_r^3} \, .
\end{eqnarray}
Here, $\ln \Lambda$ is the Coulomb logarithm. The collision frequency is then defined as:
 \begin{eqnarray}
\nu_p & = \frac{1}{p}\frac{d p}{dt} \nonumber\\
& =  \frac{1}{p} v_r \frac{d p}{d \ell} \nonumber\\
& =   n_p \frac{m_i + m_p}{m_i^2 m_p}
\frac{Z^2e^4}{4 \pi \epsilon_0^2} \frac{  \ln \Lambda}{v_r^3} \, .
 \end{eqnarray}
The effective cross section for a single plasma ion colliding with a single impurity particle is thus  given by,

\begin{equation}
\sigma(v_r) = \frac{\nu_p}{n_p v_r} =  \frac{m_i + m_p}{m_i^2 m_p}
\frac{Z^2e^4}{4 \pi \epsilon_0^2} \frac{  \ln \Lambda}{v_r^4} \, .
\end{equation}

\section{Evaluation of the integral for momentum loss rate of plasma ions}\label{appendix: integral}

Here we evaluate the integral for momentum loss rate of plasma ions to an impurity particle,

\begin{eqnarray}
\frac{\mathbf{d p}}{d t} &=& \int  (\hat{\mathbf{v}}_r \cdot (\hat{\mathbf{v}}_p - \hat{\mathbf{u}})) f_0 \sigma(v_r)v_r m_i \mathbf{v}_r \, d^3 \mathbf{v}_r\nonumber\\
&=&\int  (\hat{\mathbf{v}}_r \cdot (\hat{\mathbf{v}}_p - \hat{\mathbf{u}})) f_0 \frac{\nu_p (v_r)}{n_p} m_i \mathbf{v}_r \, d^3 \mathbf{v}_r \, .
\label{eq:dp_dt}
\end{eqnarray}
Since the dot product \( \hat{\mathbf{v}}_r \cdot (\hat{\mathbf{v}}_p - \hat{\mathbf{u}}) \) is zero when \( \hat{\mathbf{v}}_r \) is perpendicular to \( \hat{\mathbf{v}}_p - \hat{\mathbf{u}} \), evaluating the integral in equation~\ref{eq:dp_dt} leaves only the component aligned with \( \hat{\mathbf{v}}_p - \hat{\mathbf{u}} \). To take advantage of this, we align the $x$-axis of our coordinate system with \( \hat{\mathbf{v}}_p - \hat{\mathbf{u}} \), reducing the integral to the $x$-axis component and simplifying the calculation. Let us note that   this $x$-axis is used solely for calculation and is unrelated to the $x$- and $y$-axes of the domain used in the HW system in the previous section. The rate of change of momentum in the $x$-direction is given by,

\begin{eqnarray}
\frac{{d p_x}}{d t} &=&\int  \hat{v}_{r,x}  (\hat{v}_p - \hat{u}) f_0 \frac{\nu_p (v_r)}{n_p} m_i v_{r,x} \, d^3 \mathbf{v}_r\nonumber \\
&=& \frac{\nu_p(v_t)}{n_p} m_i \int \hat{v}_{r, x} (\hat{v_p} - \hat{u}) f_0 \frac{v_{r, x}}{\hat{v_r}^3}  \, d^3 \mathbf{v}_r \nonumber \\
&=& \frac{\nu_p(v_t)}{n_p} m_i (v_p - u) \int \frac{\hat{v}_{r, x}^2}{\hat{v_r}^3} f_0 \, d^3 \mathbf{v}_r \, ,
\end{eqnarray}
where the expression for $\nu_p(v_t)$ is  introduced as follows,
\begin{equation}
\nu_p(v_r) =    n_p \frac{m_i + m_p}{m_i^2 m_p}
\frac{Z^2e^4}{4 \pi \epsilon_0^2} \frac{  \ln \Lambda}{v_r^3} =  n_p \frac{m_i + m_p}{m_i^2 m_p}
\frac{Z^2e^4}{4 \pi \epsilon_0^2} \frac{  \ln \Lambda}{v_t^3 \hat{v}_{r}^3 }  = \frac{\nu_p(v_t)}{\hat{v_r}^3} \, .
\end{equation}
Then we evaluate the integral over spherical coordinates, which yields,
\begin{eqnarray}
\int \frac{\hat{v}_{r, x}^2}{\hat{v_r}^3} f_0 \, d^3 \mathbf{v}_r &=& \frac{1}{3} \int \frac{\hat{v}_{r, x}^2 + \hat{v}_{r, y}^2 + \hat{v}_{r, z}^2}{\hat{v_r}^3} f_0 \, d^3 \mathbf{v}_r \nonumber \\
&=& \frac{1}{3} \int  \frac{\hat{v_r}^2}{\hat{v_r}^3} f_0 \, d^3 \mathbf{v}_r \nonumber \\
&=& \frac{1}{3} v_t \int \frac{1}{v_r}f_0 \, d^3 \mathbf{v}_r \nonumber \\ 
&=& \frac{1}{3} v_t \int_0^\infty \frac{1}{v_r}  f_0 \, 4 \pi v_r^2 \, d v_r \nonumber \\
&=& \frac{2 \pi}{3} v_t \int_0^\infty f_0 \, 2 v_r \, d v_r \nonumber \\
&=& \frac{2 \pi}{3} v_t \frac{n_i}{(2 \pi)^{\frac{3}{2}} v_t^3} \int_0^\infty \exp \left(-\frac{v_r^2}{2v_t^2}\right) d v_r^2 \nonumber \\
&=& \frac{2 \pi}{3} \frac{n_i}{(2 \pi)^{\frac{3}{2}}} \cdot 2 = \frac{2}{3 (2 \pi)^{\frac{1}{2}}} n_i \, .
\end{eqnarray}
Thus we get 
\begin{eqnarray}
\frac{d p_x}{d t} &=& \frac{\nu_p(v_t)}{n_p} m_i (v_p - u) \int \frac{\hat{v}_{r, x}^2}{\hat{v_r}^3} f_0 \, d^3 \mathbf{v}_r \nonumber \\
&=& \frac{\nu_p(v_t)}{n_p} m_i (v_p - u) \frac{2}{3 (2 \pi)^{\frac{1}{2}}} n_i \nonumber \\
&=& \frac{m_i + m_p}{m_i^2 m_p}
\frac{Z^2e^4}{4 \pi \epsilon_0^2} \frac{  \ln \Lambda}{v_t^3} \, m_i (v_p - u) \frac{2}{3 (2 \pi)^{\frac{1}{2}}} n_i \nonumber \\
&=& n_i \frac{m_i + m_p}{m_i m_p}
\frac{Z^2e^4}{6 \sqrt{2} \pi^{3 / 2} \epsilon_0^2} \frac{  \ln \Lambda}{v_t^3} (v_p-u) \, .
\end{eqnarray}
Therefore, we have,
\begin{equation}
    \frac{d \mathbf{p}}{dt} =n_i \frac{m_i + m_p}{m_i m_p}
\frac{Z^2e^4}{6 \sqrt{2} \pi^{3 / 2} \epsilon_0^2} \frac{  \ln \Lambda}{v_t^3}  (\mathbf{v}_p - \mathbf{u}) \, . \nonumber
\end{equation}

\section*{References}
\bibliographystyle{iopart-num}
\bibliography{ppcf}

\end{document}